# Effects of fixatives on myelin molecular order probed with RP-CARS microscopy

GIUSEPPE DE VITO[1,2,3,*], PAOLA PARLANTI[1,2,4,5], ROBERTA CECCHI[1,2], STEFANO LUIN[1,6], VALENTINA CAPPELLO[2], ILARIA TONAZZINI[6], VINCENZO PIAZZA[2,7]

[1]NEST, Scuola Normale Superiore, Piazza San Silvestro 12, I-56127 Pisa, Italy
[2]Center for Nanotechnology Innovation @NEST, Istituto Italiano di Tecnologia, Piazza San Silvestro 12, I-56127 Pisa, Italy
[3]Currently with the University of Florence, Department of Neuroscience, Psychology, Drug Research and Child Health, Viale Pieraccini 6, I-50139 Florence, Italy and with European Laboratory for Non-Linear Spectroscopy, Via Nello Carrara 1, I-50019, Sesto Fiorentino, Italy
[4]The George Washington University, GW Nanofabrication and Imaging Center, 22nd Street NW, Washington, DC 20057 USA
[5]Currently with Child Health Institute of New Jersey and Department of Neuroscience and Cell Biology, Rutgers Robert Wood Johnson Medical School, 89 French Street, New Brunswick, NJ 08901 USA
[6]NEST, Istituto di Nanoscienze, Consiglio Nazionale delle Ricerche, Piazza San Silvestro 12, I-56127 Pisa, Italy
[7]Currently with Sentec Ltd - a Xylem brand, 5 The Westbrook Centre, Milton Road, Cambridge, CB4 1YG UK.
*giuseppe.devito@unifi.it

**Abstract**
When live imaging is not feasible, sample fixation allows preserving the ultrastructure of biological samples for subsequent microscopy analysis. This process could be performed with various methods, each one affecting differently the biological structure of the sample. While these alterations were well-characterized using traditional microscopy, little information is available about the effects of the fixatives on the spatial molecular orientation of the biological tissue. We tackled this issue by employing Rotating-Polarization Coherent Anti-Stokes Raman Scattering (RP-CARS) microscopy to study the effects of different fixatives on the myelin sub-micrometric molecular order and micrometric morphology. RP-CARS is a novel technique derived from CARS microscopy that allows probing spatial orientation of molecular bonds while maintaining the intrinsic chemical selectivity of CARS microscopy. By characterizing the effects of the fixation procedures, the present work represents a useful guide for the choice of the best fixation technique(s), in particular for polarisation-resolved CARS microscopy. Finally, we show that the combination of paraformaldehyde and glutaraldehyde can be effectively employed as a fixative for RP-CARS microscopy, as long as the effects on the molecular spatial distribution, here characterized, are taken into account.

**1. Introduction**
When live imaging is not feasible, the ultrastructure of biological samples must be maintained as much as possible in their native status for subsequent microscopy analysis: cellular components have to be stabilized in space and time, in a process called sample fixation. This process could be performed with physical or chemical methods, each one with its own advantages and limits [1,2].

Physical methods for biological samples fixation require the use of instruments that can be costly and challenging to be used. Among the physical approaches, cryofixation is the most important: relatively small samples (few hundreds micrometer thick) are quickly frozen to arrest instantaneously all cell activities. The main advantage of cryofixation is that this procedure leads to the ultrastructural fixation without any chemical treatment, which otherwise could introduce slight modifications of the morphology of the samples and remove its antigenicity [1,3]. Therefore, cryofixation is the best fixation method for biological samples when immunostaining experiments are required [2].

On the other hand, chemical fixation is the most commonly used method for sample preparation for light microscopy (LM) and electron microscopy (EM) analyses. In this process, fixatives stabilize macromolecules through direct chemical binding, leading to their immobilization and consequent preservation of samples architecture [4,5]. The two most used chemical fixatives are glutaraldehyde and formaldehyde. It is well known that glutaraldehyde, the most used fixative in EM, should be avoided when fluorescence imaging needs to be performed. Glutaraldehyde-fixed samples, indeed, exhibit high levels of autofluorescence when exposed to near ultraviolet light, which decreases the image contrast when other fluorescence signal is detected [2,6,7]. Conversely, formaldehyde is a much less efficient fixative and is not able, when used alone, to guarantee a satisfactory ultrastructure conservation for EM analysis [2,4].

These reasons highlight the importance of selecting the right fixative solution for the preparation of samples when specific imaging analysis has to be performed. Nevertheless, various fixation techniques affect differently the biological structure of the sample. While these alterations were well characterized using EM and LM, little information is available about the effects of the fixatives on the spatial molecular orientation of the biological tissue. In order to address this issue, we employed Rotating-Polarization Coherent Anti-Stokes Raman Scattering (RP-CARS) [8] microscopy to study the effects of different fixatives on the myelin sub-micrometric molecular order and micrometric morphology.

CARS microscopy is a multiphoton optical imaging technique, derived from Raman imaging, which allows imaging at high spatial resolution without the need to label samples or use contrast agents, being the contrast generated by the presence of selected molecular bonds [9]. While these capabilities are shared with Raman imaging, CARS allows a signal-to-noise ratio many orders of magnitude greater with respect to the latter. In the two-colour two-beam implementation of the CARS process, two incoming beams, named "pump" and "Stokes", interact in a third-order non-linear process with sample molecules to generate the detected signal [10]. If the frequency difference between the pump and Stokes beams matches the resonance frequency of molecular bonds inside the excitation volume, then the phonon population of the vibrational mode is coherently and resonantly excited, generating the optical contrast. Both Raman imaging [11] and CARS [12] were applied to study the effect of fixatives on biological samples, providing valuable information on the alteration induced by these procedures on the morphology and biochemical content of the cells and on the image contrast.

Starting from the CARS microscopy, some of us developed a new technique, named RP-CARS [8]. This modality employs a circularly-polarized Stokes beam and a linearly-polarized pump beam (with a continuously rotating polarization plane) to extract information about the orientation of the targeted molecular bonds and their degree of spatial order. This information is important since many biological tissues are characterized by a high degree of anisotropy in their molecular spatial orientation. Indeed, RP-CARS has already been proven to be an extremely useful technique to study the organization of myelin in sciatic nerves of mice [13–16] by targeting the $CH_2$ bonds. This was made possible because the structure of myelin is enriched in highly spatially-ordered $CH_2$ bonds in its lipid moiety [13], whose molecular vibrations produce a large [17,18] and polarization-sensitive [17,19] CARS signal. RP-CARS can be used to analyse both unfixed and fixed samples. Moreover, since the glutaraldehyde-induced emission at ~560 nm [7] is blue-shifted with respect to the CARS signal, the latter is not affected by its fluorescence background. Therefore, RP-CARS is the ideal tool to study comparatively the effect of fixative agents on the sub-micrometric scale.

**2. Methods**

Three wild type (WT) C57Bl/6J mice were sacrificed with cervical dislocation and sciatic nerves were rapidly extracted after their death, transversally divided in halves and fixed according to the protocols reported in Table 1. Each protocol was tested on three nerve halves. Two more mice from the same strain were used for EM acquisition and their sciatic nerves were fixed according to protocols 2 and 4. Animals were maintained under standard housing conditions and used according to the protocols and ethical guidelines approved by the Italian Ministry of Health, as per Italian law (Permit Number: 0004419- 2012 CNR, Pisa).

**Table 1. Fixation protocols employed in the study**

| | |
|---|---|
| **Protocol 1** | Unfixed sciatic nerves maintained in sodium cacodylate buffer 0.1M until the imaging was performed. Samples were observed immediately after the excision. |
| **Protocol 2** | 4 hours Para 4% in sodium cacodylate buffer 0.1M. |
| **Protocol 3** | Cryofixation with fast plunging into liquid nitrogen; samples were maintained at -80°C and unfrozen at RT right before the analysis. |
| **Protocol 4** | 4 hours Para 2% + Gluta 0.2% in sodium cacodylate buffer 0.1M. Overnight Gluta 2% in sodium cacodylate buffer 0.1M. |

We tested three different fixative protocols (2, 4: chemical fixation, 3: physical fixation), which were compared to the protocol 1 (unfixed nerve). Abbreviations: Para: paraformaldehyde; Gluta: glutaraldehyde; RT: room temperature.

The RP-CARS setup used in this work has already been described in [20]. However, for the sake of clarity, we summarize its main features and components here. Figure 1 shows a schematic of the system. The 810-nm pump beam is generated by a mode-locked Ti:Sa laser (Chameleon Vision 2; Coherent Inc.). Part of this beam, after passing through a Faraday optical isolator, is used to pump an optical-parametric-oscillator (Oria IR; Radiantis) that generates the Stokes beam at 1060 nm. The latter and the remainder of the pump beam are attenuated, expanded by telescopic beam expanders, individually chirped by SF6 optical-glass blocks and temporally overlapped by a delay line. It was shown that manipulation of the beam chirping values can lead to improvements in CARS imaging performance, such as an increase in CARS signal intensity [21] or, by exploiting the spectral focusing principle [22], an increase in spectral resolution and therefore in image signal-to-background ratio [23–25]. We pursued the latter approach: spectral focusing was achieved thanks to the optical-glass blocks by inducing a similar amount of chirping at the sample plane for the two beams [22]. As a consequence, the pulse duration is increased from ~160 fs (~115 fs) at the laser output of the pump (Stokes) beam to ~545 fs (~810 fs) on the sample [25]. Before each experimental session, the delay line position was carefully tuned, analysing the sample-emitted light with a spectrometer (iHR550, with Synapse sCCD camera, Horiba), in order to optimize the temporal overlap and the spectral selectivity for the Raman band of the $CH_2$ bonds. Before being recombined by a dichroic mirror, the polarization states of the pump and Stokes beams are conditioned by a pair of quarter-wave plates and a pair of half-wave plates (one of which is rotating by means of a gimbal motor). The Stokes beam is circularly-polarized and the pump beam is linearly-polarized with a polarization plane that continuously rotates. The overlapped beams are deflected by a pair of galvo-scanning mirrors (GVS002; Thorlabs) and routed to a high-numerical-aperture objective (C-Achroplan W; 32X, NA = 0.85, Carl Zeiss

Micro Imaging GmbH) of an inverted microscope (Axio Observer Z1; Carl Zeiss Micro-Imaging GmbH) through a scan lens, a compensation lens [20] and the tube lens of the microscope. The CARS signal is collected in the epi direction by means of a dichroic mirror, spectrally filtered to select the Raman band of the $CH_2$ bonds (2850 cm$^{-1}$) and then detected with red-sensitive photomultiplier tubes (R3896; Hamamatsu). We took at least four volumetric acquisition for each nerve segment, with an individual acquisition volume of 50 µm × 50 µm × 10 µm. The raw CARS signal is then processed by a custom-made Labview program to generate three values for each voxel: $A_{dc}$, $A_{2\omega}$ and $\varphi$, as described in [26]. $A_{dc}$ signal is similar to the traditional CARS signal, but it is free from polarization-dependent artefacts. $A_{2\omega}$ measures the average in-plane spatial anisotropy of the targeted molecular bonds in the excitation volume. Finally, $\varphi$ indicates the average in-plane orientation of the targeted molecular bonds in the same volume. From these values, using a custom-made Python program, we compute two other values: $\alpha$ and $\beta$. $\alpha$ is a quantitative and ratiometric pixel-based indicator for the molecular order described in detail in [13,14]. In order to get a single indicative value for each volumetric acquisition, we selected the voxels representing the myelin sheaths using a threshold on the $A_{dc}$ values (as described in [14]) and then we averaged their corresponding $\alpha$-values to compute the stack $\alpha$-value. In contrast, the $\beta$-value is directly computed on the volumetric stack. It corresponds to the resultant length (i.e. one minus the circular variance [27]) of the $\varphi$-values pertaining to myelin voxels and computed on pre-determined spatial length scale, as explained in [15,16] In this case, the $\beta$ parameter was calculated for 1.5 µm × 1.5 µm regions.

Protocol 1 was considered as our reference, since previous RP-CARS measurements on sciatic nerves gave very satisfying results on such unfixed samples [8,13,14]. We analysed the data using a Bayesian general linear mixed model implemented in R package "brms" [28]. We took the fixation method as fixed effect, the animal as random effect and the $\alpha$ and $\beta$ value as response variables. We employed flat functions as uninformative prior distributions.

For EM acquisitions, sciatic nerves were fixed according to protocols 2 and 4 in Table 1, then processed as previously described for resin embedding [29,30]. Briefly, samples were post-fixed with 1% osmium tetroxide and 1% potassium ferrocyanide in 0.1 M sodium cacodylate buffer, *en bloc* stained with 3% uranyl acetate in 20% ethanol, dehydrated and embedded in epoxy resin that was cured for 48 hours at 60 °C.

Samples were cut in 120 nm thin slices using a Leica UC7 ultramicrotome (Leica, Wetzlar, Germany), and sections were placed on silicon wafers (TedPellaInc., Redding, CA, USA) mounted on stubs (TedPellaInc., Redding, CA, USA) for Scanning Electron Microscopes (SEM) through carbon conductive tab (TedPellaInc., Redding, CA, USA).

The electron microscopy analysis was performed using a Helios 660 Nanofab Dual Beam FIB-SEM (ThermoFisher Scientific, Electron Microscopy Solutions, Hillsboro, OR, USA), equipped with a Concentric Backscattered Detector (CBS) which is especially indicated for high quality imaging of biological tissues sections. SEM imaging was carried out at the GW Nanofabrication and Imaging Center (GWNIC) at The George Washington University (Washington, DC, USA).

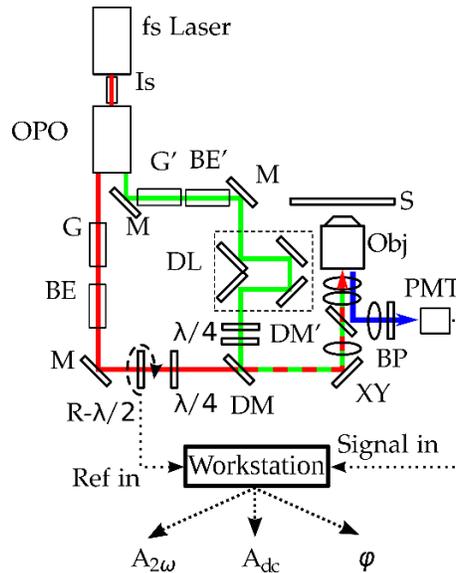

**Fig. 1.** RP-CARS setup. Femtosecond laser (fs Laser), Faraday optical isolator (Is), optical parametric oscillator (OPO), SF6 optical glasses (G and G′), beam expander (BE and BE′), dichroic mirrors (DM and DM′), rotating half-wave plate (R-λ/2), quarter-wave plates (λ/4), fixed half-wave plate (λ/2), delay line (DL), galvanometric mirrors (XY), sample (S), microscope objective (Obj), band-pass filters (BP), photomultipliers (PMT). The custom-made software (Workstation) computes $A_{dc}$, $A_{2\omega}$ and φ values from the detected signals. Thick red, green, and blue lines show the paths of the pump, Stokes and signal beams respectively. Figure and legend text adapted with permission from: "G. de Vito, A. Canta, P. Marmiroli, and V. Piazza: A large-field polarisation-resolved laser scanning microscope: applications to CARS imaging. *Journal of Microscopy*. 2015. 260. 194–199. John Wiley & Sons Ltd publisher" [20]. © 2015 The Authors. *Journal of Microscopy* published by John Wiley & Sons Ltd on behalf of Royal Microscopical Society.

## 3. Results

For the quantification of the molecular order of $CH_2$ bonds in myelin we employed the α-value metric, which describes the local orientation anisotropy of chemical bonds taking values from 0 to 90 degrees [13], where 0 indicates a completely disordered molecular system (random orientation) and 90 degrees a totally ordered system (all bonds with the same orientation) in the volume of observation, i.e. inside the sub-micrometer-sized excitation point spread function (PSF) volume. An α value around 20 is typical for nerve myelin in mice [13]. Representative results for the different fixative conditions of the myelinated nervous fibres are shown in Fig. 2 (first row), where the pixel-based α-values are mapped to a range of colours. The images were generated in the HSV (hue, saturation, value) colour-space by mapping the α-values on the H channel (from red to green), the $A_{2\omega}$ signal intensity (representative of the density of spatially ordered $CH_2$ bonds; as defined in [8]) on the S channel and the $A_{dc}$ signal intensity (analogous to traditional CARS signal) on the V channel.

From these images, it is possible to appreciate how the myelin α-values in the cryofixation condition appears higher with respect to the unfixed condition. Moreover, from the images shown in Fig. 2 it is possible to observe morphological alterations of the myelin sheath that characterize the paraformaldehyde fixation condition and the paraformaldehyde and glutaraldehyde fixation condition.

In addition to the α-value, we also visualized the φ-value: an indicator of the average in-plane local (i.e. in the excitation PSF volume) orientation of the $CH_2$ bonds. When this indicator is probed on the fibre equatorial plane from fresh tissue, it indicates the local in-plane orientation of the myelin sheath [8], due to the geometrical configuration of the $CH_2$ bonds. In order to show representative results, we colour-mapped this indicator using the HSV colour space, similarly to what we did for the α-value, by mapping it on the H channel and the results are shown in Fig. 2 (second row).

The Figure shows that in the images of the myelin sheaths appear a peculiar "punctuated" pattern with different colours exclusively for the paraformaldehyde-glutaraldehyde fixation condition.
This suggests that this chemical fixation method can induce alterations in the local average spatial orientation of the molecular bonds for this biological structure on spatial scales comparable with the excitation PSF transversal size (~0.3 μm). It is important to note that this kind of alterations are not visible in the CARS images, as they do not appear in the images constructed with the $A_{dc}$ signal (third row, analogous to traditional CARS signal); while they are evident in the RP-CARS images.

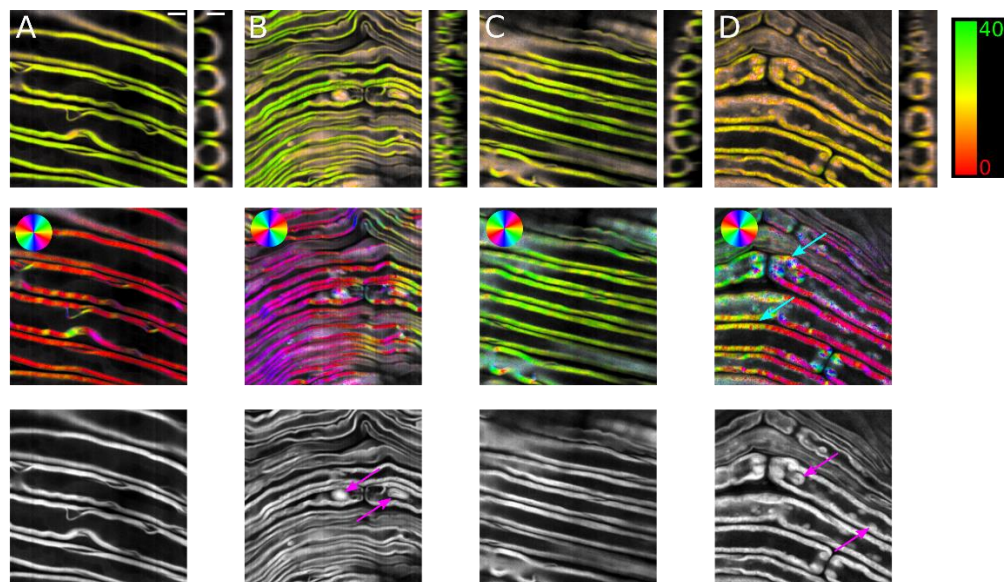

**Fig. 2.** Representative images of the different fixative conditions: unfixed (A), paraformaldehyde fixation (B), cryofixation (C), paraformaldehyde and glutaraldehyde fixation (D). First row: images generated in the HSV colour-space by mapping the α-values on the H channel, as indicated by the colour-bar on the right; the $A_{2\omega}$ signal on the S channel and the $A_{dc}$ signal on the V channel; square panels: longitudinal optical sections, side panels: transversal virtual sections. Second row: images generated in the HSV colour-space by mapping the φ-values on the H channel as indicated by the colour-wheels on the images. The S and V channels were generated as in the first row. The cyan arrows indicate examples of the punctuated colour patterns. Third row: images generated by grey-mapping the $A_{dc}$ signal. The magenta arrows indicate myelin sheath morphological alterations. Scale bar: 5 μm.

In order to quantify the observed effects, we employed a Bayesian statistical framework to analyse the results and we show the results in Figure 3. We computed a stack-based average of the α-values as described in the "Methods" section and we found that the paraformaldehyde fixation increases the average α-value (one-sided Bayes factor: ~97, posterior probability: 99%) of 2.45 (95% credibility interval: 0.83, 4.04) points. Also, the cryofixation increases the average α-value (extremely high one-sided Bayes factor, posterior probability: ~1) of 4.31 (95% credibility interval: 2.67, 5.92) points and this increase is larger compared to the paraformaldehyde fixation (one-sided Bayes factor: ~28, posterior probability: 96%). On the other hand, there is a substantial probability [31] (one-sided Bayes factor: 3.52, posterior probability: 78%) that the fixation with the combination of paraformaldehyde and glutaraldehyde slightly decreases the average α-value of 0.76 points (95% credibility interval: -2.45, 0.89).
In order to quantify the punctuated colour appearance that we observed in sample fixated with the combination of paraformaldehyde and glutaraldehyde and, in general, to quantify the effect on the φ-value distributions, we computed the stack-based β-value [16] over the spatial scale of 1.5 μm. This value varies from 0 to 1 and it indicates the orientation variance of the φ-values over image sub-regions of the chosen spatial scale and is given as the average of the β-values on all the sub-regions in which an image is divided. We choose this particular spatial scale because we observed that the differences in the β parameter among the observed sample conditions are maximized for spatial scales in the range of 1 μm - 1.5 μm and are almost constant inside this range. The β-value is linked to how frequently the orientation of the bonds change in the given length scale: when frequent changes in colour (bands or spots with a dimension comparable to the chosen spatial scale) are visible in the images generated by the φ-value colour-mapping, a low average β-value can be expected; whereas if only homogeneous colours are visible, a high average β-value can be expected.

We found a strong probability [31] that the paraformaldehyde fixation increases the average β-value (one-sided Bayes factor: ~11, posterior probability: 91%) of 0.02 (95% credibility interval: 0, 0.05) points. Also the cryofixation increases the average β-value (one-sided Bayes factor: ~240, posterior probability: ~1) of 0.05 (95% credibility interval: 0.02, 0.08) points and we found substantial probability [31] for this increase to be larger compared to the paraformaldehyde fixation (one-sided Bayes factors: ~10, posterior probability: 91%). On the other hand, we found that the fixation with the combination of paraformaldehyde and glutaraldehyde slightly decreases the average β value (one-sided Bayes factor: 119, posterior probability: 99%) of 0.04 points (95% credibility interval: -0.07, -0.01).

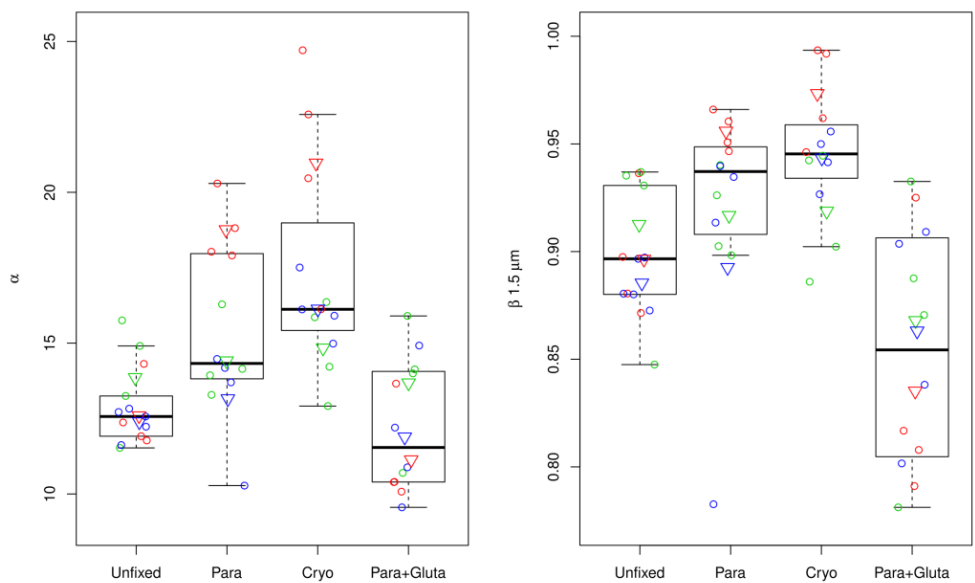

**Fig. 3.** Boxplots [32] of the observed α (left) and β (right) values with the different fixation protocols. Each hollow circle represents a microscopy volumetric acquisition. Different colours indicate different animals. The inverted triangles indicate the averages for each animal and condition. Jitter was applied on the x-axis in order to improve the readability. Abbreviations: Para: paraformaldehyde; Gluta: glutaraldehyde; Cryo: cryofixation.

Since the combination of paraformaldehyde and glutaraldehyde is widely used as fixation procedure for EM, we acquired SEM images of the myelin sheaths from murine sciatic nerves fixed either with paraformaldehyde alone or with a combination of paraformaldehyde and glutaraldehyde (following the same fixation protocols used for the CARS imaging); we show them in Fig. 4. Contrarily to the RP-CARS images, we did not find appreciable differences in the SEM images of the myelin windings between the two conditions.

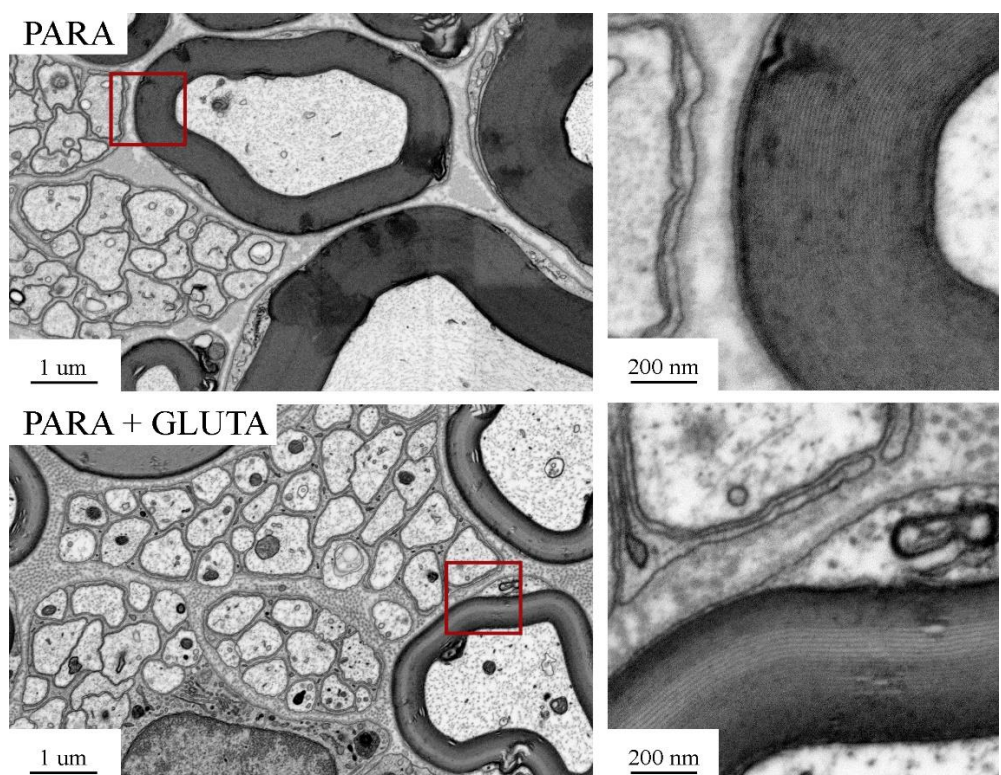

**Fig. 4.** SEM images of myelin sheaths in transverse section from murine sciatic nerves treated with two different fixation protocols, as indicated on the labels. Right column: magnifications of the areas indicated by the red rectangles. Abbreviations: Para: paraformaldehyde; Gluta: glutaraldehyde.

## 4. Discussion

We took great care in performing microscopy acquisition of unfixed samples immediately after the excision, because cells and tissues are prone to degrade and modify their structure rapidly. Therefore, we took these samples as the ground truth to be compared against the other conditions, as they represent the closest condition to the still-living tissue.

The results shown in Figs. 2 and 3 indicate that the spatial anisotropy of the $CH_2$ bonds (measured by the α-value) and their average directionality (measured by the β-value) appear to be increased in paraformaldehyde-fixed samples compared to the unfixed ones. On the other hand, the fixation procedure with the combination of paraformaldehyde and glutaraldehyde seems to alter the molecular spatial distribution of the tissue in the opposite direction, since it decreases both the parameters. This small degradation in the two parameters with respect to the unfixed sample is reasonably a consequence of the well-known tissue-shrinkage effect induced by the glutaraldehyde [33]. Moreover, the range of spatial lengths that maximizes the differences in the β-value (1 µm - 1.5 µm) corresponds to the typical dimension of the spots composing the punctuated patterns observed exclusively in this condition, thus strongly suggesting that they can be linked to the observed decrease of the β-value.

The colour-punctuated patterns are not visible with traditional CARS and this represents a strong point for the usefulness of RP-CARS and, in general, of polarisation-resolved CARS techniques. These patterns can be interpreted as micrometer-sized regions where the fixative combination altered locally the spatial orientation of the molecular spatial architecture, without however changing significantly its intrinsic structure. Moreover, when we visualized by SEM myelin sheaths fixed with paraformaldehyde alone or with a combination of paraformaldehyde and glutaraldehyde, we did not note any relevant difference. This again highlights the capabilities of RP-CARS to detect alterations in the sample molecular spatial orientation that are otherwise not easily visible with other techniques.

It should also be noted that both the paraformaldehyde alone and the combination of paraformaldehyde and glutaraldehyde induce in the tissue well-visible morphological alterations, shown in Fig. 2 (third row), which are not present in samples treated with the cryofixation procedure.

## 5. Conclusions

The presented evidences, taken together, underline that the various fixation processes, while necessary to visualize tissues for longer time and to employ several microscopy techniques, alter the molecular arrangement of the samples differently and on different spatial scales. Thus, in order to select the best technique for each application, they should be characterized, and the resultant alterations should be considered in the analysis of the obtained results. The present work contributes to this goal by observing new evidences, exploiting the capabilities of the RP-CARS microscopy technique and, in particular, its useful combination of chemical and molecular orientation sensitivity [8]. In particular, we observed that cryofixation on one hand avoids morphological alteration, but on the other it homogeneously alters the molecular spatial properties of the sample. This spatial alteration is present in a lesser degree also in samples fixated with paraformaldehyde alone, but in this case it is associated with morphological alteration too. Finally, these morphological alterations are present also when paraformaldehyde is combined with glutaraldehyde; moreover, in this case, there are also alterations in the sample molecular spatial properties, which are however different from the other cases and are not homogeneous, since they appear as punctuated patterns.

Thanks to the characterization of the effects of the fixation procedures, the present work represents also a useful guide for the choice of the right fixation technique for RP-CARS microscopy and, in general, for polarisation-resolved CARS microscopy.

Finally, in this work we show that, even if the combination of paraformaldehyde and glutaraldehyde alters the RP-CARS measure, it can be effectively employed as a fixative for this microscopy technique, as long as its characterized effects on the molecular spatial distribution are accounted for during the post-imaging analysis. As a future prospective, we found this result particularly interesting, because we could speculate further development of a new correlative microscopy between RP-CARS and EM, being the glutaraldehyde-paraformaldehyde fixation suitable for both these techniques.

## Disclosures

The authors declare no conflicts of interest.